# Maps on the basis of the *Arts & Humanities Citation Index*: the journals *Leonardo* and *Art Journal* versus "Digital Humanities" as a topic

Loet Leydesdorff [a] and Alkim Almila Akdag Salah [b]

**Abstract**

The possibilities of using the *Arts & Humanities Citation Index* (*A&HCI*) for journal mapping have not been sufficiently recognized because of the absence of a *Journal Citations Report* (*JCR*) for this database. A quasi-*JCR* for the *A&HCI* (2008) was constructed from the data contained in the Web-of-Science and is used for the evaluation of two journals as examples: *Leonardo* and *Art Journal.* The maps on the basis of the aggregated journal-journal citations within this domain can be compared with maps including references to journals in the *Science Citation Index* and *Social Science Citation Index.* Art journals are cited by (social) science journals more than by other art journals, but these journals draw upon one another in terms of their own references. This cultural impact in terms of being cited is not found when documents with a topic such as "digital humanities" are analyzed. This community of practice functions more as an intellectual organizer than a journal.

[a] Amsterdam School of Communications Research (ASCoR), University of Amsterdam; Kloveniersburgwal 48, 1012 CX Amsterdam, the Netherlands; loet@leydesdorff.net ; http://www.leydesdorff.net

[b] Virtual Knowledge Studio of the Netherlands Royal Academy of Arts and Sciences, Cruquiusweg 31, Amsterdam, the Netherlands.

## 1. Introduction

In recent decades, the bibliometric study of the sciences and the social sciences has gained more legitimacy than similar efforts to explore the humanities. However, Linmans (forthcoming) recently has argued that the humanities do not need to remain the weakest link in the scientometric enterprise. *Arts & Humanities Citation Index* (*A&HCI*) has been available since 1975, and since 2004 *Scopus* and *Google Scholar* provide alternatives which enable us to study the arts and humanities bibliometrically.

One major drawback of using the *A&HCI* for scientometric purposes has hitherto been the absence of a *Journal Citation Report (JCR)* for this database. *JCRs* are prepared annually from the *Science Citation Index* (SCI) and the *Social Science Citation Index* (SSCI) by aggregating unique citations at the journal level on an annual basis.[1] These indices enable the Institute of Scientific Information (ISI) of ThomsonReuters to compute scientometric indicators such as impact factors. *JCR*s provide scientometricians with an opportunity to develop journal maps using the matrix of aggregated journal-journal citations.

The obvious differences in the citation behavior of scholars in the arts and humanities can be considered as another drawback for applying citation analysis. Citation practices are not well established in the arts and humanities; humanities scholars rely on various media (e.g., works of art and poems) which may also be covered by the scholarly databases.

[1] *JCR*s for the *Science Citation Index* and the *Social Science Citation Index* have been published by the Institute of Scientific Information (ISI) since 1974 and 1977, respectively (Garfield, 1972, 1989).

Document types other than journal articles (e.g., books, book reviews, monographs, etc.) should be taken into account as important channels of communication in the humanities (Garfield, 1982a; Hicks & Wang, 2009; Nederhof, 2006; Nederhof & Van Raan, 1989).

Furthermore, sources in the humanities are often written by authors who gain in importance over the years. As Garfield (1979, at p. 7) formulated: "The masters are continually discussed." This "different pace of theoretical development" (Nederhof, 2006, at p. 16) results in a longer cited half-life of publications. However, differences in publication and citation practices (including cited half-lives) are pervasive among all disciplines (Leydesdorff, 2008, at p. 280). Disciplines, for example, are known to vary in terms of publication portfolio types and therefore citation patterns (Cronin *et al.*, 1997).

In a recent study of coverage of the social sciences and humanities in the bibliometric databases, Archambault *et al*. (2006) furthermore noted that a topic in the humanities or social sciences may be developed in a specific language (e.g., French) and consequently develop a semantics that is grounded in this language. In such cases, the results may be of interest mainly to local expertise. These authors also emphasized that research questions in such (sub)disciplines may not be communicable internationally, and thus perhaps less suitable for international collaboration. Such cultural factors may render research in these areas a challenge to bibliometrics, especially since the databases have often been criticized for their lack of national coverage and their overrepresentation of formal publications in the English language (e.g., Van Leeuwen *et al.*, 2001).

Despite all these shortcomings, recent years have witnessed a renewed interest in indexing and analyzing humanities research using bibliometrics (e.g., Hicks & Wang, 2009; Linmans, forthcoming). From the perspective of institutional management, administrators are under pressure to assess different faculties and departments with comparable indicators (e.g., Aksnes & Sivertsen, 2009; Butler, 2007). Is the construction of scientometric indicators and bibliometric maps using *A&HCI* a feasible and desirable project?

In 2006, the American Academy of Arts and Sciences received a three-year grant of $701,000 for developing a model set of *Humanities Indicators*. The results of this project were presented early in 2009, and are available at http://www.amacad.org/news/hrcoAnnounced.aspx. Seventy-four indicators are organized into more than 200 tables and charts. However, this data was based on a survey among scholars from the humanities; the respondents argued that indicators should focus on what "users want" instead of what the existing databases have to offer. The final chapter of the report contains some information about academic publishing in the humanities,[2] but any mentioning of citation analysis or the availability of relevant databases such as *Scopus* and the *A&HCI* remain conspicuously absent from this report.

On the other side of the Atlantic Ocean, activities to "measure the humanities" bibliometrically have been coordinated by the European Science Foundation (ESF). The ESF was established in 1974 as an association of (80+) member organizations in 30

[2] More information about this data is available at http://www.humanitiesindicators.org/content/hrcoIVD.aspx#topIV12).

European nations. In 2005, the ESF funded jointly with the European Commission a project entitled "Humanities in the European Research Area" (HERA). As a follow-up, a project for developing a "European Reference Index for the Humanities" (ERIH) was funded in 2008.

After the publication of "initial lists" ranking journals in the humanities in terms of A, B, and C-categories, the ERIH consortium announced the publication of revised lists of journals in 2009. The authors of the proposal emphasized that the distinction among the categories A, B and C was not meant to indicate quality, but to reflect factors such as the disciplinary scope and audience of a journal. The same journal can occur on several lists, but may be categorized differently depending on its importance in each discipline.[3] Nevertheless, a denunciation of the ERIH project as "dangerous" appeared in an open letter signed by more than 60 Editors of journals devoted to the history of science, technology, and medicine. These editors also demanded to have their journals removed from what they nevertheless considered as a ranking (Howard, 2008). The letter was published in the first 2009 issue of these journals, among which were leading periodicals such as *Centaurus, Perspectives on Science, ISIS, Annals of Science,* and the *British Journal for the History of Science*.[4]

Anticipating the release of the final lists by the ERIH project, Elsevier's *Scopus* announced on June 10, 2009 that it would include in its database all (approximately

[3] http://www.esf.org/research-areas/humanities/research-infrastructures-including-erih/erih-initial-lists.html
[4] This editorial can be found, for example, at http://www.publish.csiro.au/?act=view_file&file_id=HRv19n2_ED.pdf.

1,450) journals mentioned on these lists. This will bring the *Scopus* coverage of the arts and humanities to more than 3,500 journal titles.[5] The *A&HCI* currently carries approximately 1,160 journal titles.[6] The argument of the ISI is that they focus on exclusively high-quality journals. Garfield (1982b, at p. 761f.) explained that for the sake of preserving high quality, the journals in the *Science Citation Index* and *Social Science Citation Index* are "selectively screened for relevant articles in these and other fields such as history and philosophy of science, anthropology, law, economic, sociology, etc. Also, about 120 multi-authored serials, monographs, or 'books' are covered in *A&HCI*." The number of journals mentioned by Garfield (1982b, at p. 761) as fully covered source journals was 1,185 at the time.

In another context, one of us was involved in comparing the *Scopus* and ISI databases for 2007 in terms of mapping results (Leydesdorff, Moya-Anagón, and Guerrero-Bote, 2010). The *Scopus* database claimed to cover more journals in the social sciences and humanities even before the recent extension,[7] but we concluded in that context that the *A&HCI* has hitherto provided at least an equivalent resource for mapping the humanities. In this study we extend this analysis by specifying the possibilities offered by *A&HCI* in its current shape in greater detail.[8]

---

[5] http://info.scopus.com/ah/

[6] See the factsheet of the producer of the database at http://thomsonreuters.com/products_services/science/science_products/a-z/arts_humanities_citation_index. A list of all 1,480 journals ever included in the *A&HCI* can be found at http://science.thomsonreuters.com/cgi-bin/jrnlst/jlresults.cgi?PC=H.

[7] For example: "Actually Scopus includes all of the Social Sciences titles in Thomson Scientific Social Sciences Citation Index®, as well as an additional few hundred titles." at http://info.scopus.com/detail/what/julie_arnheim.asp (retrieved on August 9, 2009).

[8] The references to source items in the ISI databases are standardized more than in the *Scopus* database. Furthermore, the ISI databases reach back into the historical record by including all citations archived, while the cited references in *Scopus* go back only to 1996. Source information in the *Scopus* database older

What is specific about journals covered by *A&HCI* as potentially different from the *SCI* and *SSCI*? Do the data legitimate the decision of the ISI hitherto to refrain from producing a *JCR* for this database? How would a *JCR* of *A&HCI* inform us about these journals differently from the existing *JCRs* for the *SCI* and *SSCI*? Bollen *et al*. (2009) found a relatively higher representation of *A&H* journals in clickstream data than would be expected from citation patterns alone. Is the impact of *A&H* perhaps underestimated when focusing on the *(S)SCI* or *A&HCI*, separately?

In order to address these question, we decided to generate a quasi-*JCR* for *A&HCI* for a single year (2008) using the data at the Web of Science (WoS) and to pursue the analysis for two comparable art journals (*Leonardo* and the *Art Journal*) during the full period of their coverage both within this specific domain and by combining the three databases at the WoS interface. The results of our explorations led us to a third question—in addition to the analysis of the quasi-*JCR* for the *A&HCI* and the more fine-grained analysis of two art journals—namely, whether practices are perhaps more important in the *A&H* domain than research fields (disciplines?). We explore this question using "digital humanities" as a topic both within *A&HCI* and at the level of the three databases combined.

than 1996 is not organized systematically and cannot be used for bibliometric analysis (Ove Kähler at *Scopus*, personal communication, 27 August 2009). However, the purpose of this paper is not to compare the two databases, but to explore the different mapping options and limitations provided by the *A&HCI* in its current form. The various routines can be used with *Scopus* data, *mutatis mutandis*. However, the user should be aware that the cited references in *Scopus* data are formatted differently.

## 2. Methods and materials

One can use the citations in any set of documents downloaded from the *A&HCI* (or a combination among the three databases of the ISI) for bibliographic coupling (Kessler, 1963) or co-citation analysis (Marshakova, 1973; Small, 1973) using freely available software such as BibExcel[9] or BibJourn.[10] (A historiogram of the data can additionally be obtained using HistCite™.[11]) The download contains all the cited references in the set; the cited references contain journal names (in an abbreviated format) as a subfield. Furthermore, the recent reorganization of the ISI interface makes it possible to retrieve and download all citing documents at the so-called Citation Report of the original set. The citing documents contain again the abbreviated journal names in the cited references and thus allow one to perform a journal co-citation analysis of the set under study.

In other words, the download of all articles in a journal in a specific year generates a set containing all information about journals cited and co-cited by this journal ("citing") in that year. The "cited" pattern of the journal among other journals can comprehensively be obtained by downloading all citing documents and by repeating the analysis of the cited references in this latter set. We use the program BibJourn since it enables us to aggregate journal abbreviations in the cited references across a set of documents. The analysis can thus be pursued at the level of journals. However, the same technique can be applied to any set downloaded from the ISI databases. Such a set under study can be considered as a quasi-journal composed of documents which cite journals and which are co-cited with other journals in the references of the citing documents.

---

[9] BibExcel is freely available at http://www8.umu.se/inforsk/Bibexcel/.
[10] BibJourn is freely available at http://www.leydesdorff.net/software/bibjourn/index.htm.
[11] HistCite is available at http://www.histcite.com/ .

One can expect long tails in the distributions of journal citations. BibJourn allows for limiting the analysis, for example, to the top 1% of the citation distribution (as is default in our journal-journal routines; cf. Leydesdorff & Cozzens, 1993; Leydesdorff, 2007). In the humanities, however, we found 0.5% more appropriate because of the wide range of citations that can already be included at the top-1% level. The consequent visualizations were generated by using the algorithm of Kamada & Kawai (1989) in Pajek.[12] Citation patterns are normalized using the cosine (Ahlgren *et al*., 2003).

The size of the nodes in the figures is proportionate to the logarithm of the frequency of the citations in each network environment. In the case of using the quasi-*JCR* of the *A&HCI*—to be discussed below in more detail—the horizontal sizes are additionally adjusted to the frequency diminished with self-citations. Line-widths are proportionate to the strength of the association. A threshold of cosine > 0.2 can be used to enhance the visibility of structure in the network; this will be indicated in the legends to the figures (Egghe & Leydesdorff, 2009). The nodes are colored using the *k*-core algorithm as available in Pajek unless indicated differently.

[12] Pajek is freeware for the analysis and visualization of social networks available at http://vlado.fmf.uni-lj.si/pub/networks/pajek/.

### 2.1 Journals in the humanities

*Leonardo,* the Journal of the Society for the Arts, Sciences, and Technology, published quarterly since 1968, can be considered as a leading journal for readers interested in the applications of contemporary science and technology to the arts (Salah & Salah, 2008). This journal is not confined to the domain of the humanities, and addresses anyone following art movements which incorporate new developments in science and technology into art production. Thus, both the author profiles and audiences of the journal are oriented toward interdisciplinary exchanges. In previous reports about the *A&HCI*, Garfield (1982a, 1982b) showed that *Leonardo* was among the top-cited *A&HCI* journals during the early 80s.

We downloaded the 6,186 documents of the 41 volumes (182 issues) of the journal since 1968 (on June 8, 2009). Since the retrieved documents did not contain citation information for the period 1970-1973, the analysis was limited to publications since 1974 (5,859 documents). This set contains 31,147 cited references. As noted, using BibJourn.exe the journal names in these references can be used to construct a matrix of documents versus cited journals for each year since 1974. The resulting 35 matrices (1974-2008) were used as input to an animation using the dynamic routine created for this purpose in Visone (Leydesdorff & Schank, 2008).[13] This animation (using cosine > 0.2 as a threshold) was brought online at http://www.leydesdorff.net/journals/leonardo/citing/index.htm.

[13] The dynamic version of Visone is available at http://www.leydesdorff.net/visone .

Additionally, we use the sets of documents in each year to generate the citing sets. These documents are processed with the same methods, and the resulting animation is available at http://www.leydesdorff.net/journals/leonardo/cited/index.htm. The two animations cover all journals included which are connected to the large component in the respective set (cited or citing) at the level of cosine ≥ 0.2 in any of the years. Stress is minimized both within each year and between years using a dynamic lay-outer based on multidimensional scaling (Leydesdorff & Schank, 2008). In order to facilitate the mental map, the common nodes and links between years are kept as stable as possible during the transitions between years. Furthermore, we added substantive commentary to the animations in order to facilitate the interpretation.

| | *Number of Documents in Leonardo* (1) | *Cited References* (2) | *Number of Citing Documents* (3) | *Cited References in Citing Documents* (4) | *Ratio* (4)/(3) |
|---|---|---|---|---|---|
| 1974 | 218 | 652 | 14 | 46 | 3.3 |
| 1975 | 215 | 731 | 42 | 505 | 12.0 |
| 1976 | 256 | 763 | 60 | 664 | 11.1 |
| 1977 | 277 | 708 | 37 | 338 | 9.1 |
| 1978 | 201 | 779 | 62 | 672 | 10.8 |
| 1979 | 232 | 969 | 42 | 771 | 18.4 |
| 1980 | 235 | 1,128 | 74 | 1,105 | 14.9 |
| 1981 | 273 | 918 | 74 | 892 | 12.1 |
| 1982 | 220 | 784 | 40 | 513 | 12.8 |
| 1983 | 185 | 909 | 35 | 383 | 10.9 |
| 1984 | 158 | 799 | 25 | 708 | 28.3 |
| 1985 | 87 | 486 | 34 | 6,840 | 201.2 |
| 1986 | 98 | 684 | 53 | 8,325 | 157.1 |
| 1987 | 121 | 861 | 22 | 3,951 | 179.6 |
| 1988 | 136 | 1,174 | 41 | 4,225 | 103.0 |
| 1989 | 166 | 1,230 | 46 | 1,329 | 28.9 |
| 1990 | 131 | 1,047 | 50 | 1,163 | 23.3 |
| 1991 | 171 | 1,242 | 49 | 1,006 | 20.5 |
| 1992 | 132 | 1,192 | 50 | 1,661 | 33.2 |
| 1993 | 94 | 787 | 37 | 893 | 24.1 |
| 1994 | 127 | 935 | 39 | 1,100 | 28.2 |
| 1995 | 108 | 841 | 37 | 1,674 | 45.2 |
| 1996 | 104 | 764 | 48 | 1,593 | 33.2 |
| 1997 | 92 | 738 | 33 | 1,015 | 30.8 |
| 1998 | 124 | 1,094 | 43 | 1,575 | 36.6 |
| 1999 | 160 | 993 | 35 | 1,388 | 39.7 |
| 2000 | 223 | 851 | 30 | 931 | 31.0 |
| 2001 | 161 | 969 | 38 | 1,806 | 47.5 |
| 2002 | 176 | 1,162 | 40 | 1,307 | 32.7 |
| 2003 | 173 | 835 | 53 | 1,637 | 30.9 |
| 2004 | 162 | 848 | 47 | 1,789 | 38.1 |
| 2005 | 153 | 698 | 79 | 3,064 | 38.8 |
| 2006 | 174 | 942 | 80 | 2,756 | 34.5 |
| 2007 | 159 | 710 | 84 | 3,278 | 39.0 |
| 2008 | 157 | 924 | 107 | 4,382 | 41.0 |
| *N* | 5,859 | 31,147 | 1,680 | 65,285 | |

**Table 1**: Data used for the analysis of *Leonardo*

Table 1 provides the numbers of documents and references, citing and cited, for the years 1974-2008. The rightmost column reveals that in some years (e.g., 1985-1987) the numbers of cited references in journals citing articles published in *Leonardo* were very high. This is caused by the inclusion of bibliographies. For example, "The One Hundred Tenth Critical Bibliography of the History of Science and Its Cultural Influences," published as a special issue of *ISIS* (Neu, 1985), contained 6,269 references. In our opinion, this huge effect of bibliographies can be considered as a potential source of distortion in the citation patterns in the arts and humanities (although this effect may also occur in the *SCI* and *SSCI*).

For reasons of comparison, we selected a second journal from the *A&HCI*, namely *Art Journal.* This journal dates back to 1941, and is one of the most important journals of the *College Art Association*, an organization that can be considered the principal professional agency of the arts, art history, and art criticism in the United States. Unlike *Leonardo*, we expect this journal's referencing and being cited patterns to be more confined within the domain of the *A&HCI*, that is, less apparent at relevant interfaces between the arts & humanities and the sciences and social sciences. We limit the static comparison to the most recent year available at the time of this study, that is, to 2008. In this year, *Art Journal* published 48 papers containing 679 references. During this same year, it was cited 53 times in 42 papers (August 26, 2009).

Like most publications in the arts and humanities, both *Art Journal*'s and *Leonardo*'s citation counts are relatively low and irregular. Of the 2,645 items in *Art Journal* (in all years), 2,242 (84.8%) were never cited. These numbers are 5,011 out of 6,186 publications (81.0%) for *Leonardo*.[14] Let us therefore proceed with caution when we refer to a citation pattern of a journal in the arts and humanities. In the arts and humanities, one focuses on the tips of icebergs of possible references even more so than in the (social) sciences, since publication in the arts and humanities cannot be considered as an endogenous mechanism for generating and supporting a research front. Even if one cannot consider these maps as valid tools for evaluation purposes, they may nevertheless inform us about unexpected characteristics of these journals and reveal unforeseen aspects of the fields that support them.

### 2.2 Generation of a *Journal Citation Report* for the *A&HCI* 2008

In order to obtain a more general insight into the *A&HCI*, we constructed a *Journal Citation Report* 2008 by aggregating similarly the complete set for this one year at the journal level (Leydesdorff, 1994). This set contained 114,933 records, of which 114,929 could be retrieved, including 1,126,810 cited references based on 2,161 source journals. As noted, these journals cover approximately 1000+ sources introduced selectively into *A&HCI* in addition to the 1,157 sources that are fully covered by the *A&HCI*.[15]

[14] Ball & Tunger (2006) reported that the percentages of never cited papers older than five years in physics, mathematics, and computer science are of the order of 15, 30, and 40 percent, respectively, in the ISI database (cf. Aksnes & Sivertsen, 2004, at p. 219).

[15] Thomson Reuters lists 1,430 journals titles under the *A&HCI* at http://science.thomsonreuters.com/cgi-bin/jrnlst/jlresults.cgi?PC=H. However only 1,157 of these journal names matched records in the download for the year 2008.

A citation to an ISI source publication (in all three indices) has a standardized format as follows: "AUTHORNAME INITIALS, year, JOURNAL ABBREVIATION, Volume number, Pagenumber". Volume number and pagenumber, however, may be missing while a correct journal abbreviation is still available. Of the 1,126,810 cited references in the *A&HCI* for 2008, 1,093,005 (97.0%) contain the first three subfields of the cited reference, that is, the author-name field, the year, and the source journal or book title. These cited references were used for further processing.

Only 105,531 journal-journal citation relations (in 35,536 unique journal-journal relations) could be identified among the 2,161 journals retrieved from the *A&HCI* database in 2008. This is less than 10% of the total number of cited references (1,126,810), providing us with only a faint representation of the citation relations within this ISI domain. The other 90% of the references are to so-called "non-source publication," that is, references that are not counted as part of the *A&HCI* domain. Journal names in only 33,805 of these references (3%) match with the 8,207 journals included in the *SCI* and *SSCI*, but not in the *A&HCI*.

Only 33.0% of the remaining "source" materials in the *A&HCI* belong to the category which the ISI considers as citable issues (articles, reviews, letters, or proceedings papers), while the single category of "book reviews" takes up 42.0%. In the *Science Citation Index* 2008, 78.2% of the items are citable issues. The high percentage of book reviews supports Lewison's (2001) suggestion to use book reviews as a proxy for the impact of books. Garfield (1982b) reported on significantly similar figures for the *A&HCI* in 1981 (Table 2). The stability of this distribution ($\rho = 0.895$; $p < 0.01$) during more than 25 years is most remarkable.

| | *A&HCI 2008* | *%* | *Garfield (1982b)* | *%* |
|---|---|---|---|---|
| *Book Review* | 48,290 | 42.0 | 46,528 | 45.9 |
| *Article* | 31,883 | 27.7 | 28,210 | 27.8 |
| *Editorial Mater* | 6,803 | 5.9 | 2,217 | 2.2 |
| *Poetry* | 5,678 | 4.9 | 8,157 | 8.0 |
| *Art Exhibit Rev* | 3,361 | 2.9 | 2,457 | 2.4 |
| *Letter* | 2,513 | 2.2 | 1,529 | 1.5 |
| *Film Review* | 2,314 | 2.0 | 1,519 | 1.5 |
| *News Item* | 2,233 | 1.9 | | 0.0 |
| *Record Review* | 2,173 | 1.9 | 1,379 | 1.4 |
| *Proceedings Paper* | 1,972 | 1.7 | | 0.0 |
| *Biographical-Item* | 1,720 | 1.5 | 1,082 | 1.1 |
| *Review* | 1,665 | 1.4 | 1,481 | 1.5 |
| *Music Performan* | 1,440 | 1.3 | 1,459 | 1.4 |
| *Fiction, Creative* | 679 | 0.6 | 1,492 | 1.5 |
| *Dance Performance* | 527 | 0.5 | 579 | 0.6 |
| *Theater Review* | 505 | 0.4 | 1,581 | 1.6 |
| *Correction* | 270 | 0.2 | 76 | 0.1 |
| *Music Score Rev* | 233 | 0.2 | 755 | 0.7 |
| *TV Review, Radio* | 182 | 0.2 | 106 | 0.1 |
| *Bibliography* | 128 | 0.1 | | 0.0 |
| *Meeting Abstract* | 106 | 0.1 | 418 | 0.4 |
| *Reprint* | 92 | 0.1 | | 0.0 |
| *Excerpt* | 80 | 0.1 | 158 | 0.2 |
| *Script* | 32 | 0.0 | 124 | 0.1 |
| *Software Review* | 26 | 0.0 | | 0.0 |
| *Music Score* | 18 | 0.0 | 55 | 0.1 |
| *Database Review* | 4 | 0.0 | | 0.0 |
| *Hardware Review* | 2 | 0.0 | | 0.0 |
| Total | 114,929 | 100.0 | 101,362 | 100.0 |

**Table 2**: Document types in the *A&HCI* in 1981 and 2008.

We structured the database on the model of the *JCRs* of the ISI's other two databases. The data thus could be used directly as input to journal mapping routines already available from previous research (Leydesdorff & Cozzens, 1993; Leydesdorff, 2007). These procedures allow us to generate citation-matrices both in the cited and the citing direction, using single journals as seed journals or pre-selected journal lists. Furthermore, we can vary citation thresholds.

Although journal names are nowadays standardized in the ISI databases, a further complication arose because journal name abbreviations in the citing documents are different from journal name abbreviations in the cited references when the abbreviation contains more than 20 characters. For example, the journal *Contemporary French and Francophone Studies* is abbreviated as *Contemp Fr Francoph Stud* in the citing document, but as *Cont French Francoph* as a cited journal among the references. In order to counteract this problem, we ran a routine that assumes that if the first three words of a journal abbreviation begin with the same two characters, the journals would be considered as identical. For example, in the above example this key would be "Co-Fr-Fr".[16] The routine might introduce a bit of statistical error, but improves the number of unique journal-journal relations to 40,112 (+ 12.9%) and the total number of citation relations to 120,349 (+ 14.1%).[17]

The low numbers of standardized citations from source journals may have made the ISI hesitant to produce a *JCR* for the *A&HCI.* The matrix is extremely sparse: 36,390 of the 41,374 unique citation relations (88.0%) contain a value lower than five. For this reason, no thresholds will be used in the further analysis of matrices using this data unless otherwise specified. At http://www.leydesdorff.net/ah08/cited/index.htm and

[16] In the case of journal names with only two words, only four characters were used.
[17] The above mentioned number of 33,805 matches with journals included in the (*S*)*SCI* (but not in the *A&HCI*) was counted after this correction.

http://www.leydesdorff.net/ah08/citing/index.htm cosine-normalized citation matrices without citation thresholds in collecting the data for all 1,157 source journals to the *A&HCI* are brought online in Pajek format.

### 2.3 Digital Humanities

Journals may aggregate articles from different intellectual traditions (e.g., library *and* information sciences), but they are not by definition the most relevant units of analysis for the evaluation. New developments may take place within and/or across journals (cf. Bensman, 2007, at pp. 147 ff.; Griffith *et al*., 1974; Small, 1978). If the journal is not primarily a unit of intellectual organization in the arts and humanities, but mainly a channel of cultural dissemination, may intellectual exchange then be organized topically? We shall turn to the topic of "digital humanities" and show the possibilities of the proposed methods of mapping when applied either to a set extracted from the *A&HCI* itself or in combination with the two other databases at the *WoS* interface.

The "digital humanities," previously known as "humanities computing," can be considered as a community of practice (Agyris and Schön 1978). The topic itself is defined and applied differently by practitioners with a variety of disciplinary backgrounds. For example, it can be considered as a tool or methodology enabling humanities research, teaching, presentation, and preservation methods.[18] Academic departments that make use of digital humanities laboratories typically include technical practitioners as well as traditionally trained scholars. Such departments tend to be heavily involved in collaborative and interdisciplinary research projects with colleagues in other departments.[19]

[18] See Wiliam McCarty's (2003) essay "Humanities Computing" for an evaluation of the term and a conceptualization of the workflow among different disciplines and their methodologies in Digital Humanities research.

| | *Number of Documents* | *Cited References* | *Number of Citing Documents* | *Cited References in Citing Documents* |
|---|---|---|---|---|
| *A&HCI* | 23 | 429 | 24 | 2,181 |
| *WoS* | 46 | 829 | 30 | 2,480 |

**Table 3**: data about "digital humanities" (since 1975)

Despite the proclaimed priority of this topic as the part of an envisaged cyber-infrastructure relevant for the humanities (e.g., at http://www.neh.gov/ODH/GrantOpportunities/tabid/57/Default.aspx), the search string 'ts=("digital humanities" or "humanities computing")' in the *WoS* generated only two documents in the *A&HCI* 2008 and two more in the other two databases. For this reason, we extended in this case the search to all years (since 1975). This provided us with 23 and 46 documents, respectively (on September 8, 2009; Table 3).[20] We use the aggregated cited references in and citations to these documents for the mapping.

[19] This definition of "digital humanities" is based partly on the one at http://en.wikipedia.org/wiki/Digital_humanities, retrieved on September 17, 2009.
[20] The search tag "ts" stands for topical search. This search retrieves documents with matches in title words, abstract words or keywords attributed to the document. A search with the search terms in the title ("ti") retrieved in this case 14 and 22 documents, respectively.

## 3. Results

### 3.1 The journal *Leonardo*

The animations—at http://www.leydesdorff.net/journals/leonardo/cited/index.htm and http://www.leydesdorff.net/journals/leonardo/citing/index.htm, respectively—locate *Leonardo* as an interdisciplinary journal connected to the sciences, social sciences, and the arts throughout the time span covered. Note that the journal started its publication with an interdisciplinary intention and orientation. This interdisciplinarity in its citation environment did not change over the years in terms of either its referenced knowledge base ("citing") or its ("cited") impact environment. However, the citation patterns are not dense and are therefore volatile from year to year.

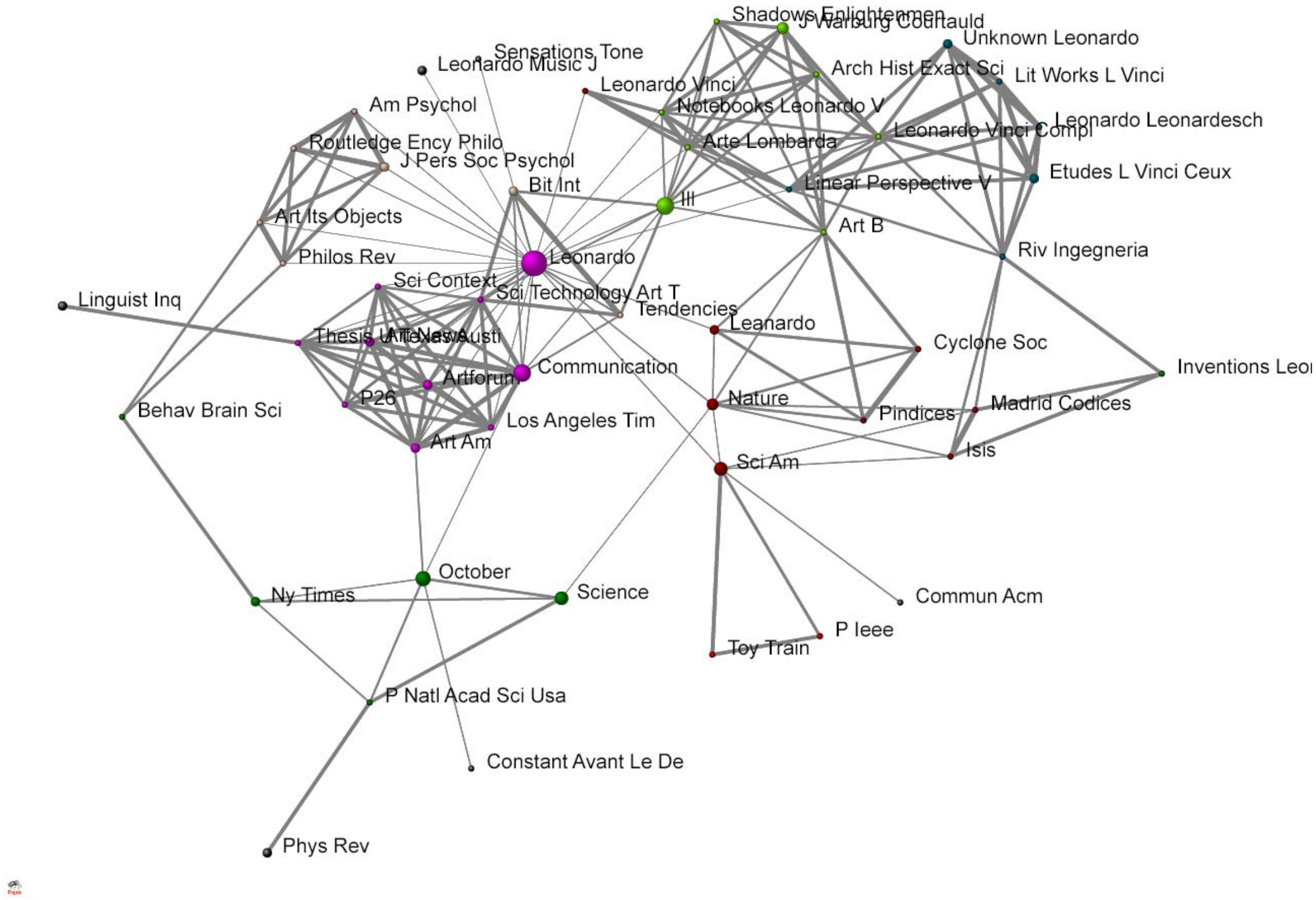

**Figure 1:** 53 journals cited by 157 articles in *Leonardo* in 2008; no citation threshold within the set; cosine > 0.0.

Figure 1 shows the results of the co-citations of 53 journals cited in 924 references of the 157 articles published in *Leonardo* during 2008.[21] Figure 2 provides the corresponding co-citation map using 107 articles which cited *Leonardo* in 2008. While articles in *Leonardo* cite the sciences and the social sciences in addition to its citations to journals and books in the humanities, the journal is mainly cited in domains other than the arts and humanities given the threshold of using only journals which contribute 0.5% to the aggregate of the references.

[21] Five more journals were cited, but not co-cited with any of the journals in the set.

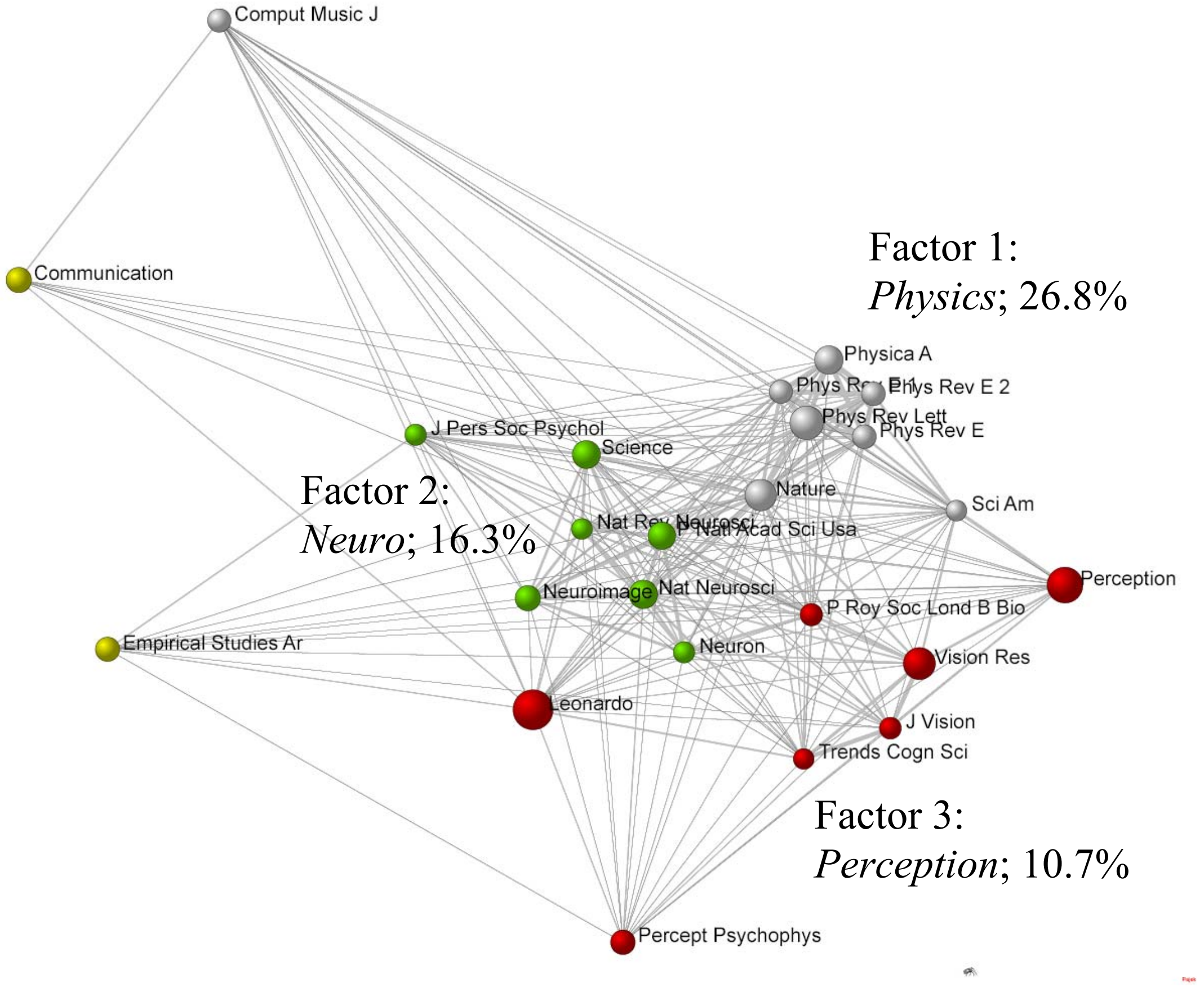

**Figure 2**: Cosine relations among 3,259 references in 107 articles citing *Leonardo* during 2008; only journals which contribute more than 0.5% to the total number of citations; no citation threshold within the set; cosine > 0.0; colors of nodes correspond to the highest factor loadings in a varimax-rotated three-factor solution.

Figure 2 shows an unexpected finding: among the 568 journals which contain documents citing *Leonardo* in 2008 (3,259 times), only 24 contribute more than 0.5% (that is, more than 16 times) to its being cited pattern, and these 24 journals are mainly in the domain of the sciences and the social sciences. Three journal groups are relevant in this citation impact environment: physics, neuroscience, and perception research. These three factors explain 53.8% of the variance. (The nodes are colored in accordance to their highest loadings in the Varimax-rotated three-factor solution.) *Leonardo* itself is positioned at the edge between the latter two specialties.

The presence of science journals in this being-cited environment of *Leonardo* is not stable over the years, but in all years science journals are visible in relatively large clusters. Among the science journal citing *Leonardo* from year to year, *Science, Nature,* and *Scientific American* dominate the animation. Another cluster contains journals with a focus on computer graphics since the mid-80s. A third, relatively stable cluster is provided by journals in cognitive science that enter the picture at the beginning of the 1990s, with strong connections to a psychology cluster. Through studies on vision and perception, journals in neuroscience, cognitive science, psychology, and computer graphics are related to this citation environment.

Art journals form another important group in the animations of the citation patterns of *Leonardo*. This group is more or less equally divided into journals focusing on aesthetics, art theory, and contemporary art news. Upon closer inspection of the animations, one can distinguish that theoretically oriented journals are more persistently visible in the animation based on articles citing *Leonardo*, while journals reporting on the latest state of the art market cite *Leonardo* more than they are cited in *Leonardo*'s publications. Core journals of the arts and art history such as *Art News*, *Studio International*, *Art Forum, Art Bulletin,* and *Art Journal* are consistently included.

After the turn of the century, *Leonardo* has increasingly lost citations from the art world in favor of citations from journals in the sciences. In 2008 (Figure 2), science journals are prevailing in its citation impact environment. In earlier years, however, some core-books by Gombrich (1960), Arnheim (1954 and 1969), and Goodman (1988) were also heavily cited. These art historians are renowned for their interest in psychology and linguistics, and hence their presence as references in the citation networks strengthened *Leonardo*'s citation relations with journals in these disciplines.

b. *Leonardo* in the quasi-*JCR* data of the *A&HCI*

Using the quasi-*JCR* of the *A&HCI*, one focuses exclusively on the source journals included in the *A&HCI*, whereas all cited references were included in the analysis when using the *WoS* interface. Within the more restricted domain of the *A&HCI*, however, only 44 journals were cited by articles in *Leonardo* during 2008; 36 of these form a citation network at the level of cosine > 0.2, as shown in Figure 3.

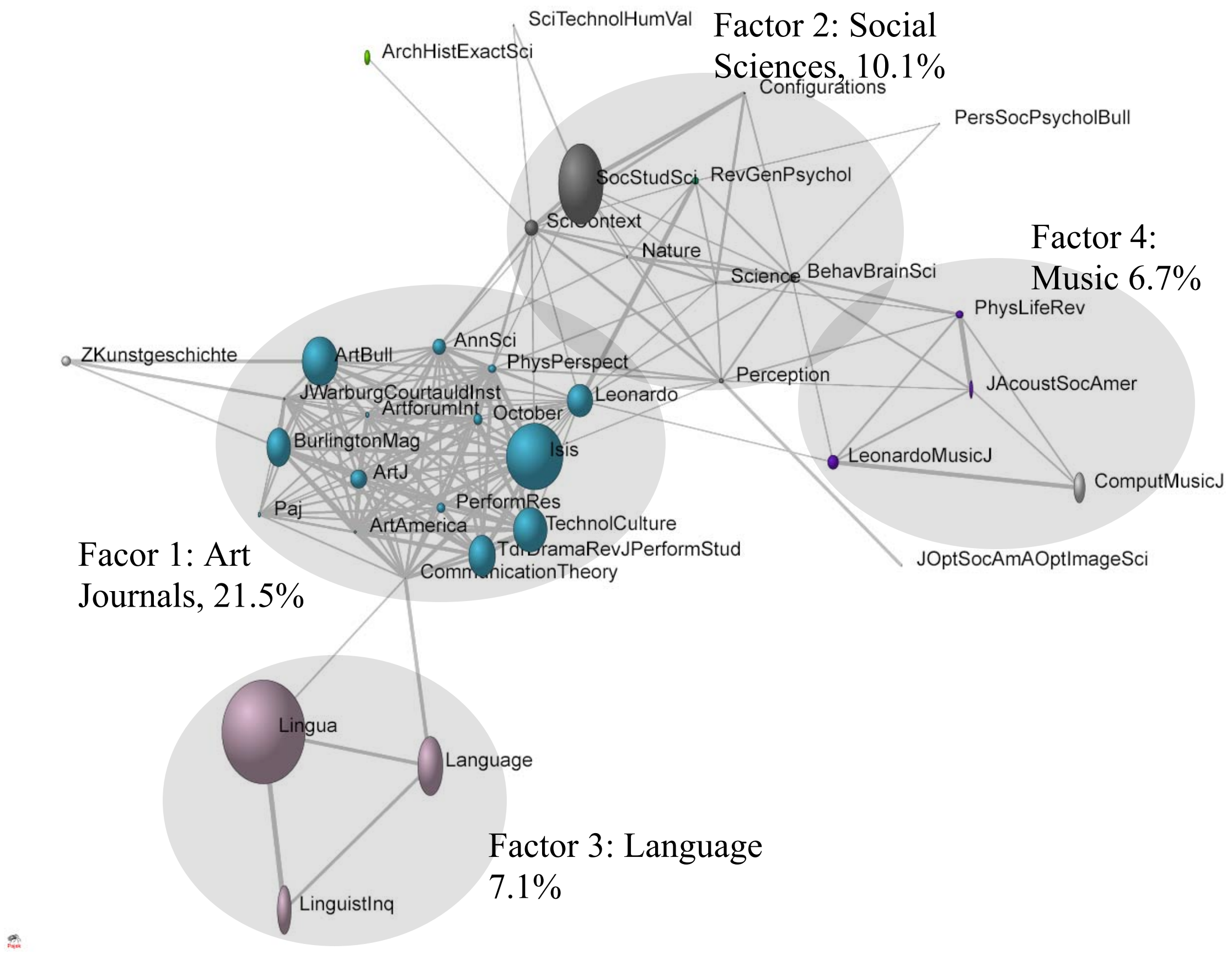

**Figure 3:** Journal map of 36 journals in the *A&HCI* cited by publications in *Leonardo* 2008; no citation threshold; cosine ≥ 0.2.[22]

*Leonardo* is positioned in this representation among other art journals such as *Art Bulletin, Artforum International,* the *Burlington Magazine,* and *Art America.* The journal's interdisciplinary position is no longer visible given this domain of exclusively the *A&HCI*. However, journals in the history and philosophy of science (e.g., *ISIS, Social Studies of Science*) form a circle at a first distance ($k$ = 1).

[22] Three more journals were not connected to the largest component at the level of cosine ≥ 0.2.

The factor analysis distinguishes four factors which are designated with their respective percentages of explained variance in Figure 3. The social sciences, linguistics, and computer music are distinguishable as separate groups. (We return to the *Zeitschrift für Kunstgeschichte* which is visible on the left side of the picture, in a section below.)

We can conclude that the art world itself is a relatively closed group in terms of referencing, revealing only weak links to the sciences and linguistics group. The comparison of Figures 1 and 3 shows that *Leonardo* draws citing upon a varied knowledge base, but is more central to its environment when the three databases are combined (Figure 1). The citing environment in the *A&HCI* (Figure 3), however, enables us to delineate a cluster of neighboring journals that provide an intellectually organized knowledge base referenced by publications in *Leonardo*.

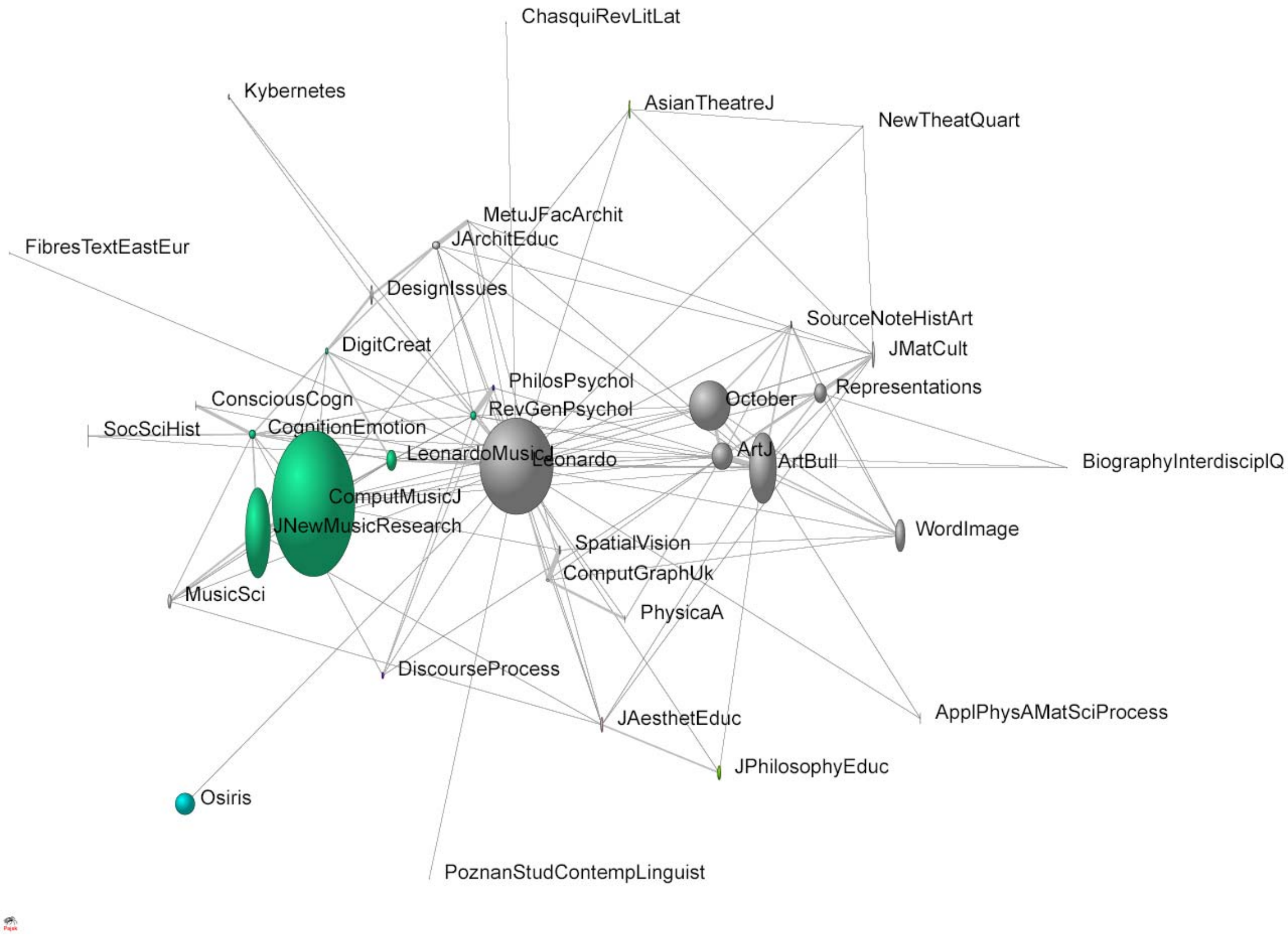

**Figure 4:** 36 journals in the citation impact environment of *Leonardo* in the *A&HCI* 2008; no citation thresholds; cosine > 0.0.

The citation impact environment of *Leonardo* in the restricted domain of the *A&HCI* 2008 (Figure 4) is almost barren in comparison to Figure 2, where three strong clusters in the science literature were visible. Here, *Leonardo* is mainly embedded in a cluster of art journals. The rest of the network can be considered as an environment to this cluster: the remaining journals are loosely connected to one another and to the main component. Among them, a few cognitive science journals and a computer graphics journal catch the eye. Other than these journals, Figures 2 and 4 share no resemblances at all, as Figure 2 did not contain a single art journal. This discrepancy in *Leonardo*'s citation impact as assumed by using the *WoS* or the quasi-*JCR* of the *A&HCI* as relevant environments raises the question of whether this diffusion pattern is specific for *Leonardo* or more generally the case for art journals.

c. *Art Journal*

The editors of *Leonardo* stated the objective of creating an interdisciplinary publication venue. Our findings show that the journal has indeed generated an interdisciplinary citation impact environment. In the citing dimension we were able to identify a group to which the journal can be attributed, but only after confining the analysis to the *A&HCI* domain. Would such a pattern be the norm for journals in the arts and humanities? Might other art journals also have a more general impact on the basis of cultural values, that is, beyond lines of intellectual organization, while drawing their references (that is, citing) from a specific knowledge base? Let us compare the citation patterns of *Leonardo* with a second art journal.

*Art Bulletin* is the first journal that comes to mind for such a comparison. This leading art journal has been published since 1913 by the *College Art Association*, the main organization of art historians in the USA. *Art Bulletin* had higher rankings than *Leonardo* in Garfield's (1982b) most-cited journal lists of *A&HCI*; in 2008, it had 283 citations in this database as against 90 for *Leonardo.* However, as a long-standing journal dedicated to publishing the latest research in the history of art and architecture, the publication scope and audience of *Art Bulletin* is not close enough to *Leonardo* for a fair comparison. *Art Bulletin* is devoted to scholarly discussions and emphasizes theory and methodology of art history, while *Leonardo* is focused on contemporary art, and covers art events and related news in addition to discussions of theory and methods.

For this reason, we chose to use another publication of the *College Art Association*, namely *Art Journal*. This journal publishes (since 1941) articles related to contemporary art, and in that sense its audience and constituency is akin to that of *Leonardo*. In the quasi-*JCR* of the *A&HCI* 2008, *Art Journal* is cited 86 times by articles in 49 journals. One hundred sixty-two journals, however, cite papers from *Art Journal* during 2008 in the larger domain of the combined *SCI, SSCI,* and *A&HCI*; 161 of these 162 journals are related to the main component above the level of cosine > 0.2.

In other words, *Art Journal* like Leonardo is overwhelmingly cited outside the domain of the arts and the humanities. The journals in this larger environment range from physics to advertising research, but most references are to "non-source" journals such as the *NY Times, Newsweek,* and the *Washington Post.* In other words, even more than in the case of *Leonardo*, the citation impact of *Art Journal* shows a large network of influence, and a large imbalance between being cited and citing. The impact of journals in the arts is not confined to the *Arts & Humanities* as scholarly discourses in journals, but reaches a much wider set including the sciences, the social sciences, and the larger public.

Using the restricted set of the *A&HCI* it is possible to select a set of 27 journals which are associated to *Art Journal* and to one another as a core, with 24 relations among each two of them at the level of cosine > 0.2. Both *Leonardo* and the *Art Bulletin* are part of this core set (Figure 5).

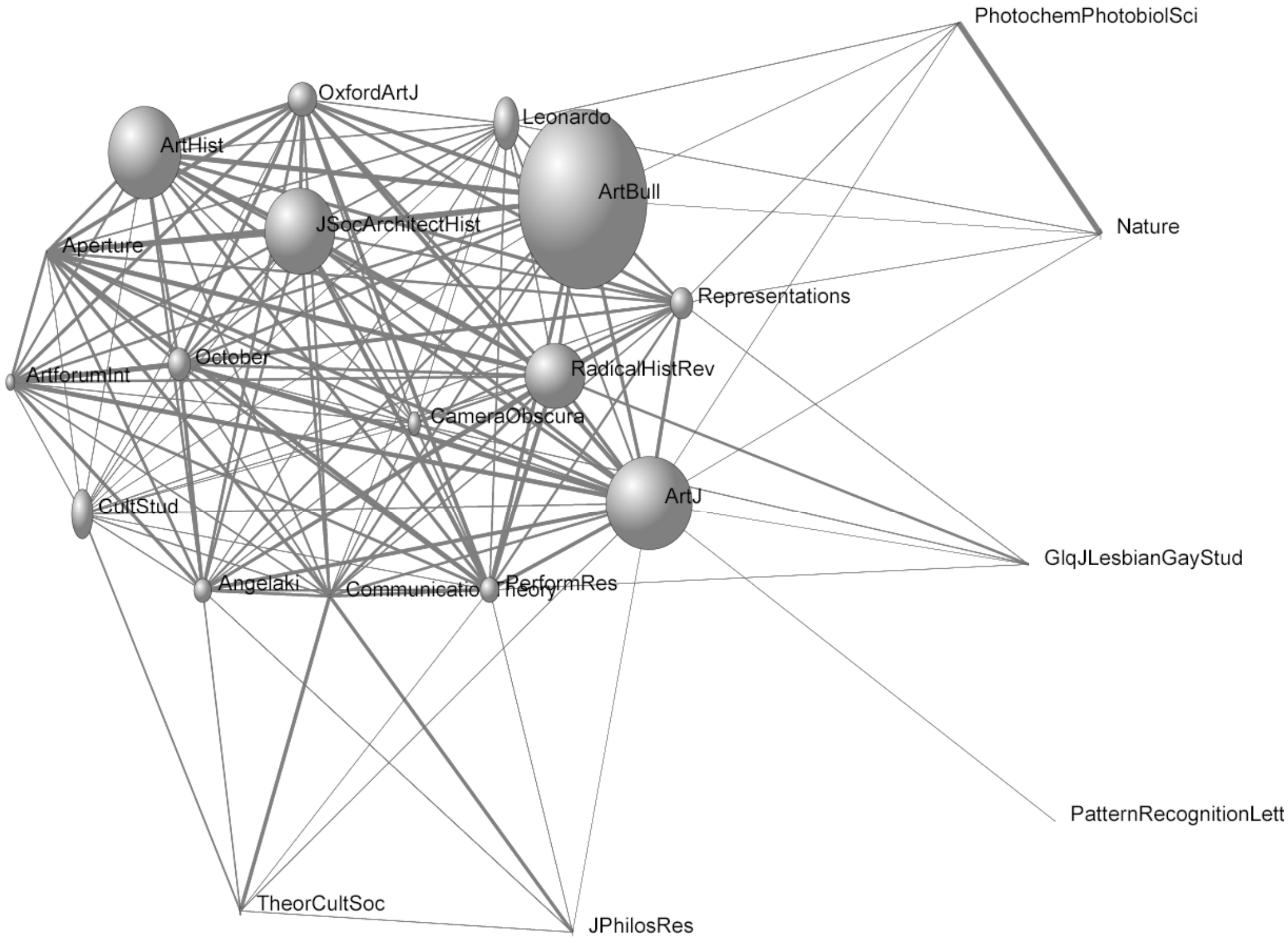

**Figure 5**: 22 journals cited by *Art Journal* in the *A&HCI* domain in 2008; no citation threshold; cosine > 0.0.

In summary, the patterns of citations in the citing and cited dimensions are different for these art journals. Although they draw on a wider environment, it is possible to find core groups among the journals in the *A&HCI* in terms of how the authors in these journals provide references when constructing their arguments. These journals, however, are not cited primarily in these restricted environments, but in the larger environment, perhaps not so much for intellectual as for cultural and instrumental reasons. The predominant rationale of references to these journals is different from that which governs the sciences and the social sciences, where intellectual organization can explain the patterns of citation.

Given this conclusion, one might indeed be hesitant to assess journals and research covered by the *A&HCI* in terms of scientometric indicators which use field-specific parameters. These journals may have functions completely different from the specialty structures that prevail in the sciences and social sciences. Thus, the journals and the constituting articles can be evaluated also in terms of these wider cultural influences. The database and citations are organized not only on socio-cognitive grounds, but also on the basis of cultural patterns.

### 3.3 Languages as cultural organizers

The priority of cultural patterns of diffusion brings us back to Archambault *et al.*'s (2006) argument that these alternative paths of cultural dissemination might be local, regional, or national. Let us pause for a moment with this possibility. Table 4 first shows the language distribution among the 114,929 items retrieved for the purpose of the construction of a quasi-*JCR* of the *A&HCI* 2008.

| | *Frequency* | *Percent* |
|---|---|---|
| English | 82,298 | 71.6 |
| French | 12,520 | 10.9 |
| German | 8,691 | 7.6 |
| Italian | 3,669 | 3.2 |
| Spanish | 3,344 | 2.9 |
| Russian | 1,613 | 1.4 |
| Czech | 532 | 0.5 |
| Dutch | 489 | 0.4 |
| Croatian | 296 | 0.3 |
| Chinese | 231 | 0.2 |
| Polish | 196 | 0.2 |
| Lithuanian | 191 | 0.2 |
| Slovene | 179 | 0.2 |
| Turkish | 149 | 0.1 |
| Slovak | 138 | 0.1 |
| Portuguese | 122 | 0.1 |
| Swedish | 102 | 0.1 |
| *other* | 169 | 0.1 |
| *Total* | 114,929 | 100.0 |

**Table 4**: language distribution of the items contained in the *A&HCI* 2008.

The shares for the major European languages are higher than in the *Science Citation Index* (see a comparable table for the *SCI* 2007 and *Scopus* 2007 in Leydesdorff *et al*., 2010, Table 1). It is noteworthy that there are no contributions in Chinese, and that only four documents are in Japanese. However, our investigations left us nevertheless with the impression that the citation patterns were not organized primarily along linguistic lines.

For example, Figure 6 shows the cited impact environment for the journal *Zeitschrift für Kunstgeschichte* in the quasi-*JCR.* The journal is cited in 2008 by articles in 23 journals of the *A&HCI*. Figure 6 is based on the matrix of citations among these journals. Unlike the *Zeitschrift für Soziologie* (see Figures 11 and 12 in Leydesdorff *et al*., 2010), this journal is cited in other art journals at the international level: German journals are interfaced with journals in other European languages.

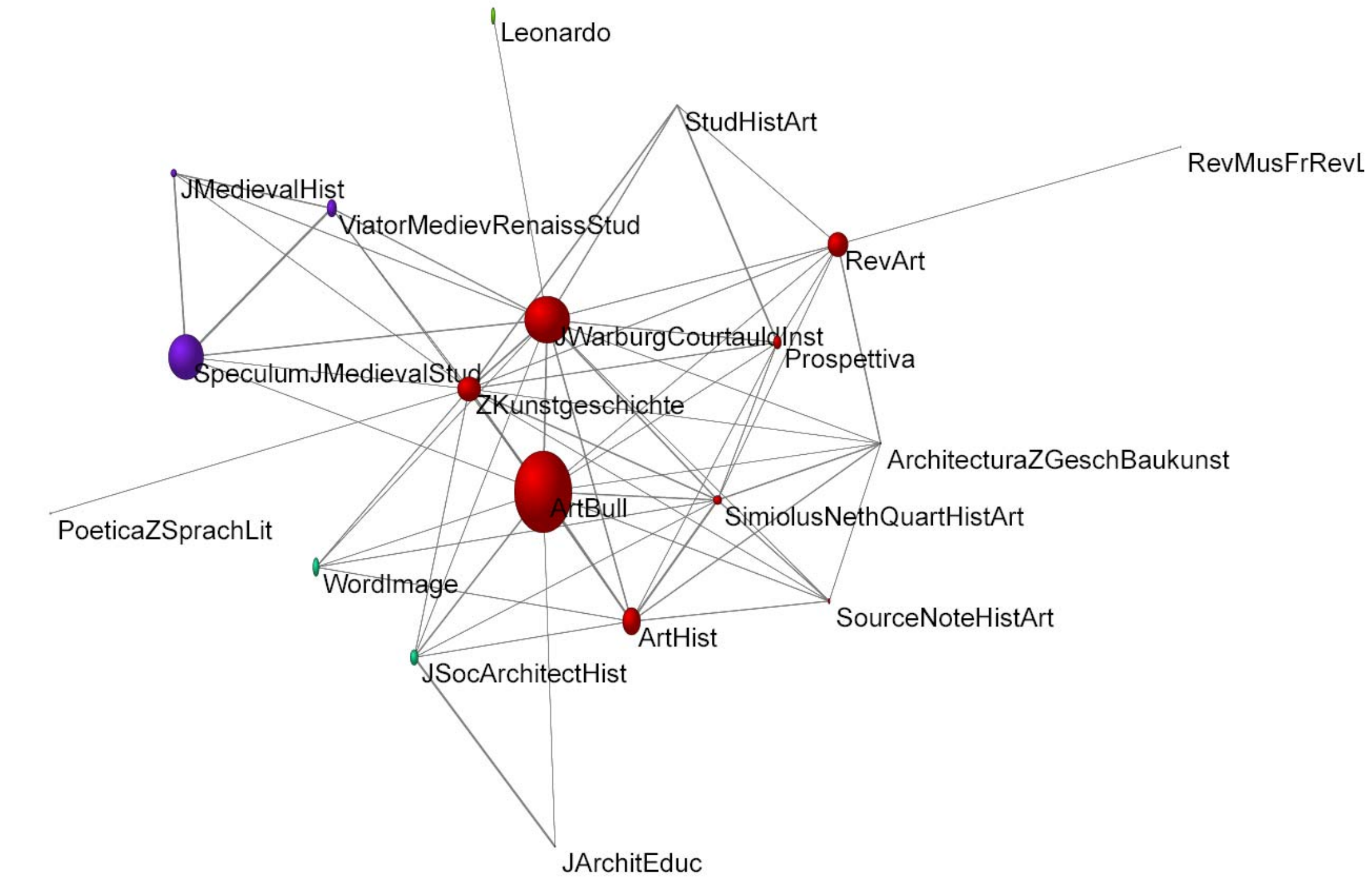

**Figure 6**: Network of 21 journals citing articles in the *Zeitschrift fuer Kunstgeschichte* in the *A&HCI* 2008; no citation threshold; cosine > 0.2.[23]

Although art history and therefore the *Zeitschrift für Kunstgeschichte* are deeply embedded in German scholarly traditions, this journal's citation pattern is more internationalized than that of comparable journals in the social sciences. Art journals may be embedded internationally more than social science journals because cultural transmission within the West increasingly transcends the boundaries of nations.

[23] Two more journals did not relate at the level of cosine ≥ 0.2.

### 3.4 Digital Humanities

If journals in the arts and humanities serve functions other than intellectual organization, this begs for the question of whether other mechanisms of organization and intervention such as professional communities can replace journals in providing the function of intellectual focus and exchange. Let us for this purpose explore the initiatives in what is nowadays mostly called "digital humanities," but was previously known as "humanities computing."

Digital Humanities can be considered as a focus or topic that was first generated by new possibilities to apply computational analysis methods to large collections, as well as to archive works of arts and literature in a digital format. Depending on the research environment, Digital Humanities can be envisioned as a methodology, a tool, or a research initiative to investigate how knowledge is produced with new media technologies, while at the same time making use of these technologies in humanities research itself. As noted, one can also consider Digital Humanities as a community of practice.[24]

[24] See Raben (1998) "Humanities computing 25 years later," Hockney (2004) "The history of humanities computing" for an overview of the praxis, and McCarty (1998), Unsworth (2002), Busa (2004) and Piez (2008) for discussions on how digital humanities should be applied.

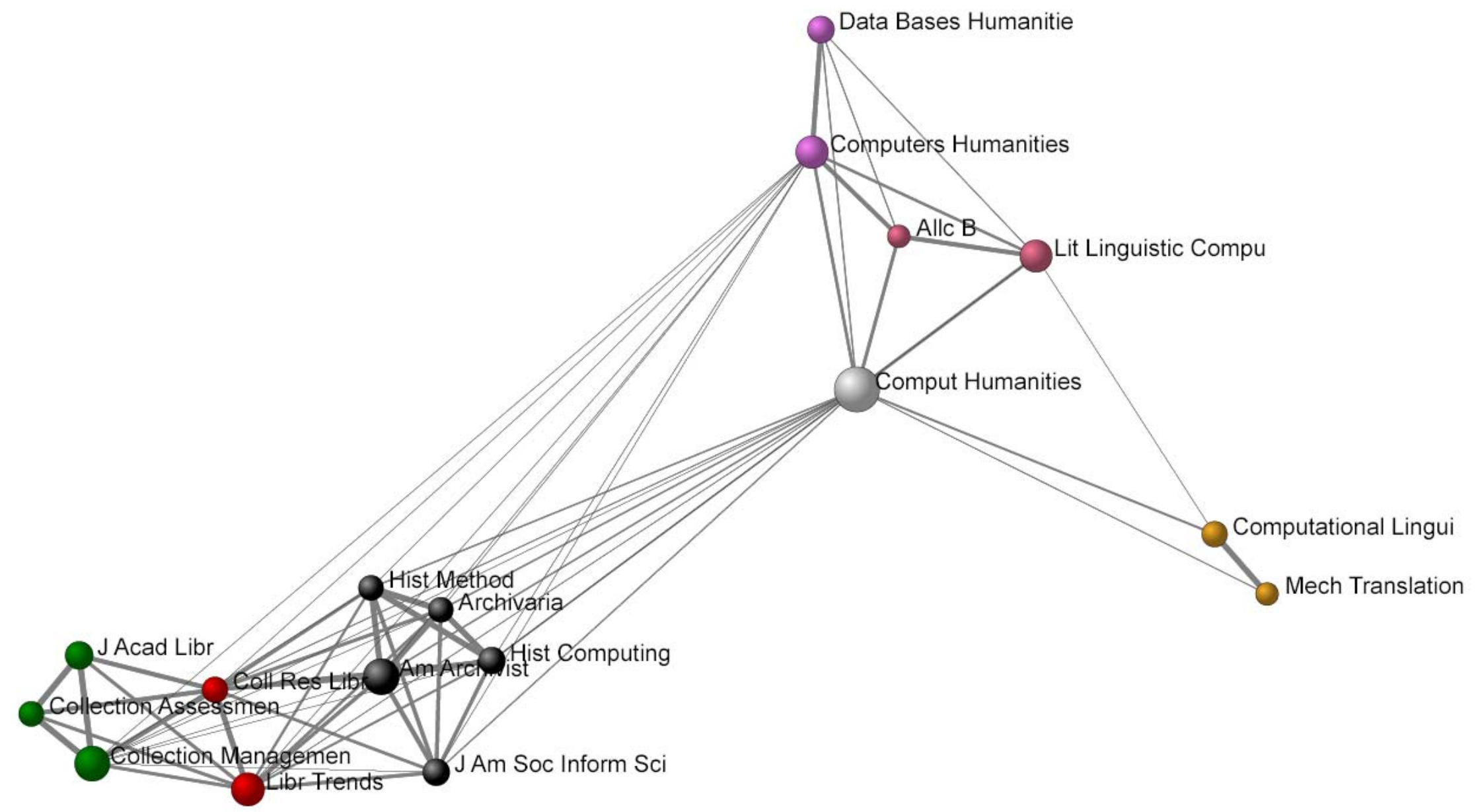

**Figure 7**: Journal co-citation patterns of 33 documents citing 46 documents about "Digital Humanities" in 2008; threshold 0.5%; cosine ≥ 0.0.

Figures 7 provides the citation impact environment for the 46 documents downloaded under "digital humanities" or "humanities computing" from the ISI's *Web of Science* on September 8, 2009. The figure shows that these documents are cited in a limited domain of two or three groups of journals, namely, new specialist journals with a focus on computer usage in the humanities, and a group of library and information science journals addressing the digitalization of archives and libraries.

These 46 documents cite from a wider range of disciplines including 81 journals. Six factors explain 66.5% of the variance in this matrix. The first three factors (51.7%) can be designated, respectively, as library & information science (34.6%), the application of computers in linguistics (10.9%), computers and literature, including markup languages (6.1%). Factor 1 can be considered as a strong component of 30+ journals. This core group is made visible in Figure 8.

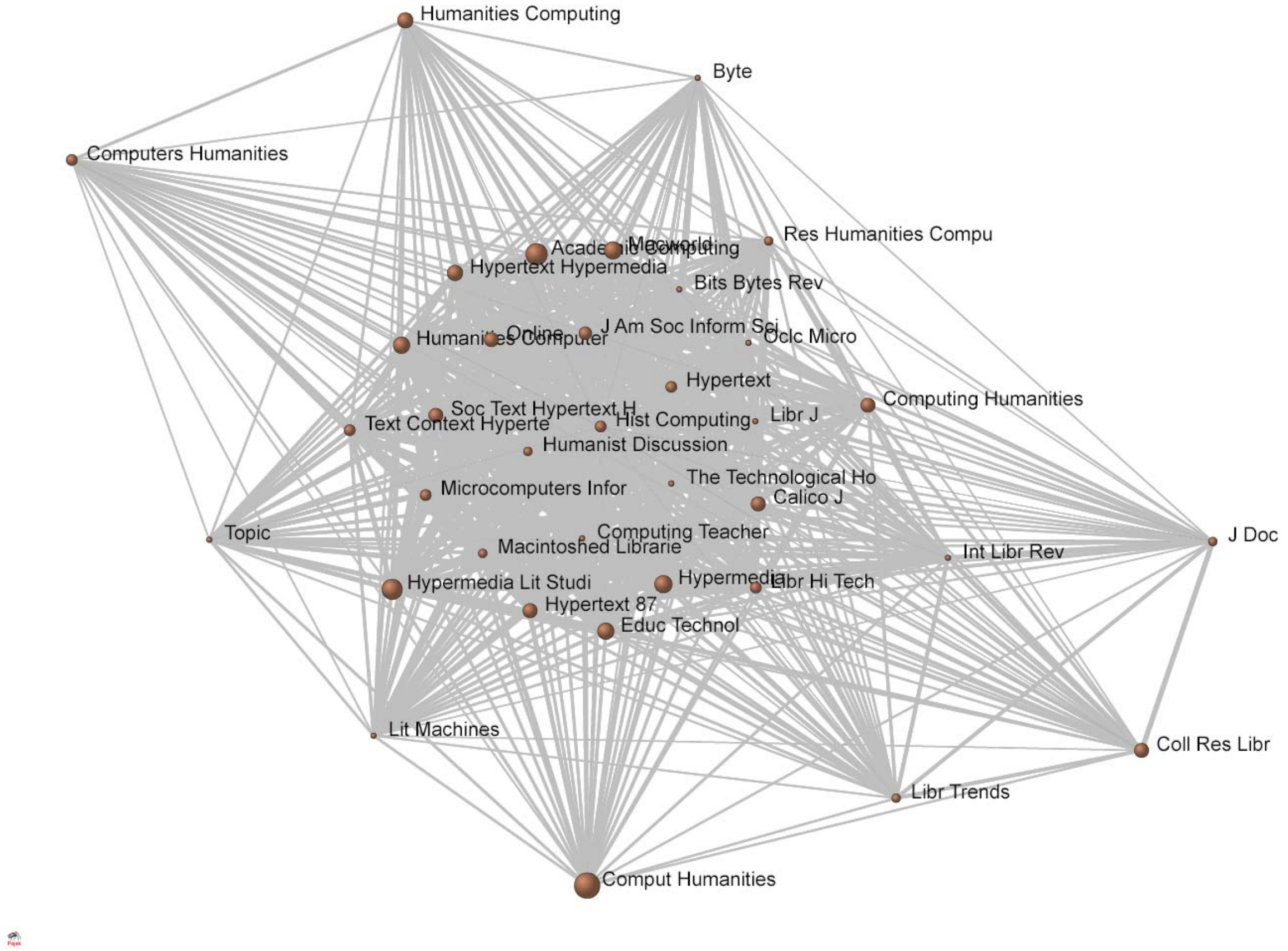

**Figure 8**: *k*-core group of 36 journals bibliographically coupled in 46 documents about "digital humanities"; threshold 0.5%; cosine > 0.2.

The majority of the journals in this core group have a long publication history that goes back to the first decade of "humanities computing." Journals like *Academic Computing*, *Humanities Computing*, or *CALICA Journal* were important publication venues for scholars combining computational methods with humanities research, long before "digital humanities" was formulated as a topic. In addition to such determining journals for the Digital Humanities, one encounters journals such as *Byte*, a computer art journal with a publication history of 30 years, or *Macworld*, a magazine with articles about the latest technological achievements of Apple industries. These journals are cited not only because of the scholarly activities they report, but also because of the need to share information about the latest technologies, or about how these technologies can be applied in settings beyond those for which they were originally designed (such as using early computers to create art).

Unlike the humanities journals that we investigated in previous sections, "digital humanities" as a topic does not provide us with a wide-spread pattern in its citation impact environment. Its citing and cited patterns resemble rather more those of the (social) sciences, in the sense that the impact is limited to a few groups of scholarly journals. Among these groups computational linguistics and text analysis are central. Interestingly enough, information visualization—a topic that is currently one of the main occupations in laboratories developing Digital Humanities—is represented neither in the citing nor the cited maps.

In summary, the investigation of this dataset retrieved from the *WoS* reveals that the topic does not mirror the citation patterns of journals in the humanities, but is more akin to that in the other sciences. In terms of citation flows, furthermore, the topic is not so much diffused into humanities as one would expect, but equally related to disciplines such as linguistics and computer science.

In our opinion, the numbers of documents for the whole period 1975-2008 were astonishingly low. As noted, Digital Humanities can be considered as a community of practice(s) more than a specialty. Following the advent of the Internet, Digital Humanities scholars made use of this new venue both for doing research and for publishing and sharing information. However, the dataset collected from the *WoS* represents only the formal literature and therefore disregards most communications that appear in online journals, discussion forums, blogs, mailing lists, etc. To map this larger knowledge base of the Digital Humanities, a dataset including these venues (e.g., Google Scholar) could be considered, but such an elaboration would reach beyond the scope of this study.

## Conclusions and discussion

Given the absence of a *JCR* for the *A&HCI*, we reconstructed journals in terms of the citing and being-cited patterns using the user interface of the ISI databases at the *Web of Science*. Additionally, we aggregated the complete set of documents attributed to the *A&HCI* for the construction of a quasi-*JCR* in 2008. Our first interest was in the position in their citation networks of typical art journals like *Leonardo* and *Art Journal.* Although these two journals address different audiences, namely an interdisciplinary one in the case of *Leonardo* and one more focused on expertise in the arts in the case of *Art Journal*, we found similar patterns in both cases. The two journals are widely cited beyond their "disciplinary" background. We proposed to consider this cultural dissemination as different from the intellectual organization that prevails in the sciences and the social sciences.

Using the restricted set of the quasi-*JCR* for the *A&HCI* 2008, it is possible to retrieve a cluster of journals in both cases that use references from other art journals in a comparable way. Thus, one could say that these journals belong to an intellectually coherent group in their reference patterns, but not in their citation patterns. Since evaluation studies measure impact by being cited, this raises questions for the evaluation of these journals using scientometric indicators (e.g., impact factors). Impact in the arts may mean something different from the sciences and the social sciences.

Our results suggest that the being-cited patterns in these cases do not indicate the provision of a knowledge base for new knowledge contributions at a research front, but may mean a source of cultural inspiration and influence. This would also explain the slower pace of "progress" in the humanities. We showed that the being-cited pattern of a community of practice (the "digitial humanities") was more focused in this domain than that of these journals. The concern among scholars and journal editors in the arts and humanities about the administrative tendency to evaluate the arts and humanities using indicators like those applied in the sciences and the social sciences should thus be taken seriously.